\documentclass[aps,prd,twocolumn,groupedaddress,amssymb,eqsecnum,showpacs,epsfig]{revtex4}
\usepackage{graphicx}
\usepackage{bm}
\usepackage{dcolumn}
\usepackage{amsmath}
\def \lleq {\lower0.9ex\hbox{ $\buildrel < \over \sim$} ~}
\def \ggeq {\lower0.9ex\hbox{ $\buildrel > \over \sim$} ~}

\def \ob  {\Omega_b}
\def \obh {\Omega_b h^2}
\def \om   {\Omega_{\rm 0m}}

\def \beq  {\begin{equation}}
\def \eeq  {\end{equation}}
\def \ber  {\begin{eqnarray}}
\def \eer  {\end{eqnarray}}

\begin{document}
\newcommand{\newc}{\newcommand}

\newcommand{\ben}{\begin{eqnarray}}
\newcommand{\een}{\end{eqnarray}}
\newc{\be}{\begin{equation}}
\newc{\ee}{\end{equation}}
\newc{\ba}{\begin{eqnarray}}
\newc{\ea}{\end{eqnarray}}
\newc{\bea}{\begin{eqnarray*}}
\newc{\eea}{\end{eqnarray*}}
\newc{\D}{\partial}
\newc{\ie}{{\it i.e.} }
\newc{\eg}{{\it e.g.} }
\newc{\etc}{{\it etc.} }
\newc{\etal}{{\it et al.}}
\newcommand{\nn}{\nonumber}
\newc{\ra}{\rightarrow}
\newc{\lra}{\leftrightarrow}
\newc{\lsim}{\buildrel{<}\over{\sim}}
\newc{\gsim}{\buildrel{>}\over{\sim}}
\title{Comparison of Standard Ruler and Standard Candle \\constraints on Dark Energy Models}
\author{R. Lazkoz$^a$, S. Nesseris$^b$, and L. Perivolaropoulos$^b$}
\email{http://leandros.physics.uoi.gr} \affiliation{$^a$Fisika
Teorikoa, Zientzia eta Teknologiaren Fakultatea, Euskal Herriko
Unibertsitatea, 644 Posta Kutxatila, 48080 Bilbao, Spain\\
$^b$Department of Physics, University of
Ioannina, Greece}
\date{\today}

\begin{abstract}
We compare the dark energy model constraints obtained by using
recent standard ruler data (Baryon Acoustic Oscillations (BAO) at
$z=0.2$ and $z=0.35$ and Cosmic Microwave Background (CMB) shift
parameters $R$ and $l_a$) with the corresponding constraints
obtained by using recent Type Ia Supernovae (SnIa) standard candle
data (ESSENCE+SNLS+HST from astro-ph/0701510). We find that, even
though both classes of data are consistent with $\Lambda$CDM at the
$2\sigma$ level, there is a systematic difference between the two
classes of data. In particular, we find that for practically all
values of the parameters $(\Omega_{\rm 0m},\ob)$ in the $2\sigma$
range of the the 3-year WMAP data (WMAP3) best fit, $\Lambda$CDM is
significantly more consistent with the SnIa data than with the
CMB+BAO data. For example for $(\om,\ob)=(0.24,0.042)$ corresponding
to the best fit values of WMAP3, the dark energy equation of state
parametrization $w(z)=w_0 + w_1\frac{z}{1+z}$ best fit is at a
$0.5\sigma$ distance from $\Lambda$CDM $(w_0=-1,w_1=0)$ using the
SnIa data and $1.7\sigma$ away from $\Lambda$CDM using the CMB+BAO
data. There is a similar trend in the earlier data (SNLS vs CMB+BAO
at $z=0.35$). This trend is such that the standard ruler CMB+BAO
data show a mild preference for crossing of the phantom divide line
$w=-1$, while the recent SnIa data favor $\Lambda$CDM. Despite of
this mild difference in trends, we find no statistically significant
evidence for violation of the cosmic distance duality relation $\eta
\equiv \frac{d_L(z)}{d_A(z) (1+z)^2}=1$. For example, using a prior
of $\om=0.24$, we find $\eta=0.95 \pm 0.025$ in the redshift range
$0<z<2$, which is consistent with distance duality at the $2\sigma$
level.
\end{abstract}

%
%
\maketitle

\section{Introduction}
The accelerated expansion of the universe has been confirmed during
the last decade by several observational probes
\cite{Riess:2004nr,Spergel:2006hy,Readhead:2004gy,Goldstein:2002gf,Rebolo:2004vp,Tegmark:2003ud,Hawkins:2002sg}.
The origin of this acceleration may be attributed to either dark
energy with negative pressure, or to a modification of General
relativity that makes gravity repulsive at recent times on
cosmological scales. In order to distinguish between these two
possibilities and identify in detail the gravitational properties of
dark energy or modified gravity two developments are
required\cite{Boisseau:2000pr}:
\begin{enumerate}
\item Detailed observation of linear cosmic density perturbations
$\delta (z)=\frac{\delta \rho(z)}{\rho}$ at recent redshifts. \item
Detailed mapping of the expansion rate $H(z)$ as a function of the
redshift $z$.
\end{enumerate}
The later is equivalent to identifying the function $w(z)$ defined
as \be w(z)\,=\,-1\,+\frac{1}{3}(1+z)\frac{d\ln (\delta
H(z)^2)}{d\ln z}\,,{\label{wzh1}} \ee where $\delta
H(z)^2=H(z)^2/H_0^2-\Omega_{\rm 0m} (1+z)^3-\Omega_{0r} (1+z)^4$
accounts for all terms in the Friedmann equation not related to
matter and radiation. If the origin of the accelerating expansion is
dark energy then $w(z)$ may be identified with the dark energy
equation of state parameter $w(z)=\frac{p_X}{\rho_X}$. The
cosmological constant ($w(z)=-1$) corresponds to a constant dark
energy density.

It has been shown \cite{Vikman:2004dc} that a $w(z)$ observed to
cross the line $w(z)=-1$ (phantom divide line) is very hard to
accommodate in a consistent theory in the context of General
Relativity. On the other hand, such a crossing can be easily
accommodated in the context of extensions of General Relativity
\cite{Boisseau:2000pr}. Therefore, the crossing of the phantom
divide line $w=-1$ could be interpreted as a hint in the direction
of modified gravity. Such a hint would clearly need to be verified
by observations of linear density perturbation evolution through eg
weak lensing \cite{Refregier:2006vt} or the redshift distortion
factor \cite{Hamilton:1997zq}.

There are two classes of probes that may be used to observe the
expansion rate $H(z)$ or equivalently $w(z)$ \begin{itemize} \item
{\it Standard Candles} are luminous sources of known intrinsic
luminosity which may be used to measure the luminosity distance
which, assuming flatness, is connected to $H(z)$ as \be d_L (z)= c
(1+z) \int_0^z dz'\frac{1}{H(z')} \label{dl1} \ee Useful standard
candles in cosmology are Type Ia supernovae
\cite{Riess:2004nr,Riess:2006fw,Davis:2007na,Astier:2005qq} (SnIa)
and the less accurate but more luminous Gamma Ray Bursts
\cite{Meszaros:2006rc} \item {\it Standard Rulers} are objects of
known comoving size which may be used to measure the angular
diameter distance which, in a flat universe, is related to $H(z)$ as
\be d_A (z)= \frac{c}{1+z} \int_0^z dz'\frac{1}{H(z')} \label{da1}.
\ee The most useful standard ruler in cosmology is the last
scattering horizon, the scale of which can be measured either
directly at $z\simeq 1089$ through the CMB temperature power
spectrum or indirectly through Baryon Acoustic Oscillations (BAO) on
the matter power spectrum at low redshifts. Clusters of
galaxies\cite{Bonamente:2005ct,Uzan:2004my} and radio
galaxies\cite{Daly:2007pp} may also be used as standard rulers under
certain assumptions but they are less accurate than CMB+BAO.
\end{itemize}

Early SnIa data put together with more recent such data through the
Gold dataset \cite{Riess:2004nr,Riess:2006fw} have been used to
reconstruct $w(z)$, and have demonstrated a mild preference for a
$w(z)$ that crossed the phantom divide line
\cite{Nesseris:2005ur,Nesseris:2006er}. A cosmological constant
remained consistent but only at the $2\sigma$ level. However, the
Gold dataset has been shown to suffer from systematics due to the
inhomogeneous origin of the data \cite{Nesseris:2006ey}. More recent
SnIa data (SNLS\cite{Astier:2005qq}, ESSENCE\cite{WoodVasey:2007jb},
HST\cite{Riess:2006fw}) re-compiled in \cite{Davis:2007na} have
demonstrated a higher level of consistency with $\Lambda$CDM and
showed no trend for a redshift dependent equation of state.

On the other hand, the use of standard rulers (CMB+BAO) has rarely
been studied independent of SnIa due to the small number of
datapoints involved (see however
\cite{Corasaniti:2007rf,Alam:2006kj,Wang:2007mza}). It has been
pointed out\cite{Percival:2007yw} that the latest BAO data ``require
slightly stronger cosmological acceleration at low redshifts than
$\Lambda$CDM ''. This statement is equivalent to a trend towards a
$w(z)<-1$ at low $z$, and therefore a possibility of crossing the
PDL $w=-1$. The goal of this paper is to quantify this statement in
some detail by comparing the best fit form of $w(z)$ obtained from
the SnIa data to the corresponding form obtained from the CMB+BAO
data. This comparison is done quantitatively by identifying the
quality of fit of $\Lambda$CDM in the context of each dataset. In
particular, we consider the Chevalier-Polarski-Linder (CPL)
\cite{Linder:2002dt,Chevallier:2000qy} parametrization \be
w(z)=w_0+w_1\frac{z}{1+z} \label{cpl} \ee and, assuming flatness, we
identify the ``distance'' in units of $\sigma$ ($\sigma$-distance)
of the parameter space point $(w_0,w_1)=(-1,0)$ corresponding to
$\Lambda$CDM from the best fit point $(w_0,w_1)$ for each dataset
(SnIa standard candles or CMB+BAO standard rulers) and for several
priors of $(\Omega_{\rm 0m},\ob)$. We thus identify an interesting
systematic difference in trends between the two datasets.

We also discuss the implications of this difference in trends on the
{\it distance duality relation} \be \eta(z)\equiv
\frac{d_L(z)}{d_A(z) (1+z)^2}=1 \label{distdual} \ee which measures
quantitatively the agreement between luminosity and angular diameter
distances. This relation has been shown to be respected when
clusters of galaxies are used as standard rulers \cite{Uzan:2004my}.

\section{Likelihood Calculations}
We assume a CPL parametrization for $w(z)$ and apply the maximum
likelihood method separately for standard rulers (CMB+BAO) and
standard candles (SnIa) assuming flatness. The corresponding late
time form of $H(z)$ for the CPL parametrization is \ba H^2
(z)&=&H_0^2 [ \Omega_{\rm 0m} (1+z)^3 + \nn
\\ &+& (1-\Omega_{\rm 0m})(1+z)^{3(1+w_0+w_1)}e^{\frac{-3w_1 z}{(1+z)}}]
\label{hcpl} \ea At earlier times this needs to be generalized
taking into account radiation ie \be E^2(a)\equiv
\frac{H(a)^2}{H_0^2}=\Omega_m(a+a_{eq})a^{-4}+ \Omega_{de}{X(a)} \ee
where $a=1/(1+z)$, $\Omega_{de}=1-\Omega_m-\Omega_{rad}$ and \ben
X(a)&=&{\rm exp}\left[-3\int_{1}^a\! \frac{(1+w(a'))}{a'}da'\right]
\nn \\ &=& a^{-3(1+w_0+w_1)}e^{-3w_1 (1-a)} \een with the CPL
parametrization $w(a)=w_0 + w_1 (1-a)$.

\subsection{Standard Rulers}
\subsubsection{CMB}
We use the datapoints $(R,l_a,\obh)$ of Ref. \cite{Wang:2007mza}
where $R$, $l_a$ are two shift parameters: \begin{itemize} \item The
scaled distance to recombination \be R=\sqrt{\om \frac{H_0^2}{c^2}}
\; r(z_{CMB}) \label{shiftR} \ee where $r(z_{CMB})$ is the comoving
distance from the observer to redshift $z$ and is given by \be
r(z)=\frac{c}{H_0}\int_0^z\frac{dz}{E(z)} \ee with $E(z)=H(z)/H_0$.
\item The angular scale of the sound horizon at recombination \be
l_a=\pi \frac{r_(a_{CMB})}{r_s(a_{CMB})} \label{shiftl} \ee where
$r_s(a_{CMB})$ is the comoving sound horizon at recombination given
by \be
r_s(a_{CMB})=\frac{c}{H_0}\int_0^{a_{CMB}}\frac{c_s(a)}{a^2E(a)} da
\label{rs}\ee with the sound speed being $c_s(a)=1/\sqrt{3(1+\bar
{R}_b a )}$ and $a_{CMB}=\frac{1}{1+z_{CMB}}$, where $z_{CMB}=1089$.
Actually, $z_{CMB}$ has a weak dependence on $\Omega_m$ and
$\Omega_b$ \cite{Hu:2000ti} but we have checked that the sound
horizon changes only to less than $0.1 \%$. The quantity $\bar
{R}_b$, is actually the photon-baryon energy-density ratio, and its
value can be calculated using $\bar {R_b}=\frac{3}{4}\frac{\Omega_b
h^2}{\Omega_{\gamma} h^2}= 31500 \Omega_b h^2 (T_{CMB}/2.7 K)^{-4}$.
\end{itemize}
For a flat prior, the 3-year WMAP data (WMAP3) \cite{Spergel:2006hy}
measured best fit values are \cite{Wang:2007mza}
\begin{eqnarray}
\bf{{\bar V}_{CMB}} &=& \left(\begin{array}{c}
{\bar R} \\
{\bar l_a}\\
{\bar \obh}\end{array}
  \right)=
  \left(\begin{array}{c}
1.70\pm 0.03 \\
302.2 \pm 1.2\\
0.022 \pm 0.00082 \end{array}
  \right)
\label{cmbdat} \end{eqnarray} The corresponding normalized
covariance matrix is \cite{Wang:2007mza}
\begin{eqnarray}
 {\bf C^{norm}_{CMB}}=\left(
\begin{array}{ccc}
 1 & -0.09047 & -0.01970 \\
 -0.09047 & 1 & -0.6283 \\
 -0.01970 & -0.6283 & 1
\end{array}
\right)
\end{eqnarray}
from which the covariance matrix can be found to be: \be (
C_{CMB})_{ij}=(C^{norm}_{CMB})_{ij} ~ \sigma_{{\bar V}^i_{CMB}}
\sigma_{{\bar V}^j_{CMB}} \ee where $\sigma_{{\bar V}^i_{CMB}}$ are
the $1\sigma$ errors of the measured best fit values of eq.
(\ref{cmbdat}).

We thus use equations (\ref{cmbdat}), (\ref{shiftR}) and
(\ref{shiftl}) to define
\begin{eqnarray}
\bf{X_{CMB}} &=& \left(\begin{array}{c}
R - 1.70 \\
l_a-302.2\\
\Omega_bh^2-0.022\end{array}
  \right),
\end{eqnarray}
and construct the contribution of CMB to the $\chi^2$ as
\be
\chi^2_{CMB}=\bf{X_{CMB}}^{T}{\bf C_{CMB}}^{-1}\bf{X_{CMB}} \ee with
\begin{eqnarray}
 {\bf C_{CMB}}^{-1}=\left(
\begin{array}{ccc}
 1131.32 & 4.8061 & 5234.42\\
 4.8061 & 1.1678 & 1077.22 \\
 5234.42 & 1077.22 & 2.48145\times 10^6
\end{array}
\right),
\end{eqnarray}
Notice that $\chi^2_{CMB}$ depends on four parameters ($\om$,
$\ob$, $w_0$ and $w_1$). Due to the large number of parameters
involved, in what follows we will consider various different
priors on the parameters $\om$, $\ob$.

\subsubsection{BAO}
As in the case of the CMB, we apply the maximum likelihood method
using the datapoints \cite{Percival:2007yw}

\begin{eqnarray}
\bf{{\bar V}_{BAO}} &=& \left(\begin{array}{c}
\frac{r_s(z_{CMB})}{{D_V(0.2)}} =0.1980 \pm 0.0058  \\
\frac{r_s(z_{CMB})}{{D_V(0.35)}} =0.1094 \pm 0.0033 \end{array}
  \right),
\end{eqnarray}
where the dilation scale \be D_V
(z_{BAO})=\left[\left(\int_0^{z_{BAO}} \frac{dz}{H(z)}\right)^2
\frac{z_{BAO}}{H(z_{BAO})}\right]^{1/3} \label{dv1} \ee encodes the
visual distortion of a spherical object due to the non-Euclidianity
of a FRW spacetime, and is equivalent to the geometric mean of the
distortion along the line of sight and  two orthogonal directions.
We thus construct
\begin{eqnarray}
\bf{X_{BAO}} &=& \left(\begin{array}{c}
\frac{r_s(z_{\rm dec})}{{D_V(0.2)}}  - 0.1980 \\
\frac{r_s(z_{\rm dec})}{{D_V(0.35)} }- 0.1094 \end{array} \right),
\end{eqnarray}
and using the inverse covariance matrix \cite{Percival:2007yw}
\begin{eqnarray}
{\bf C_{BAO}}^{-1} &=& \left(\begin{array}{cc}
     35059 & -24031 \\
    -24031 & 108300\end{array}
  \right),
\end{eqnarray}
we find the contribution of BAO to $\chi^2$ as \be
\chi^2_{BAO}=\bf{X_{BAO}}^{T}{\bf C_{BAO}}^{-1}\bf{X_{BAO}} \ee
\begin{figure*}[!t]
\begin{minipage}{18pc}
\hspace{0pt}\rotatebox{0}{\resizebox{1\textwidth}{!}{\includegraphics{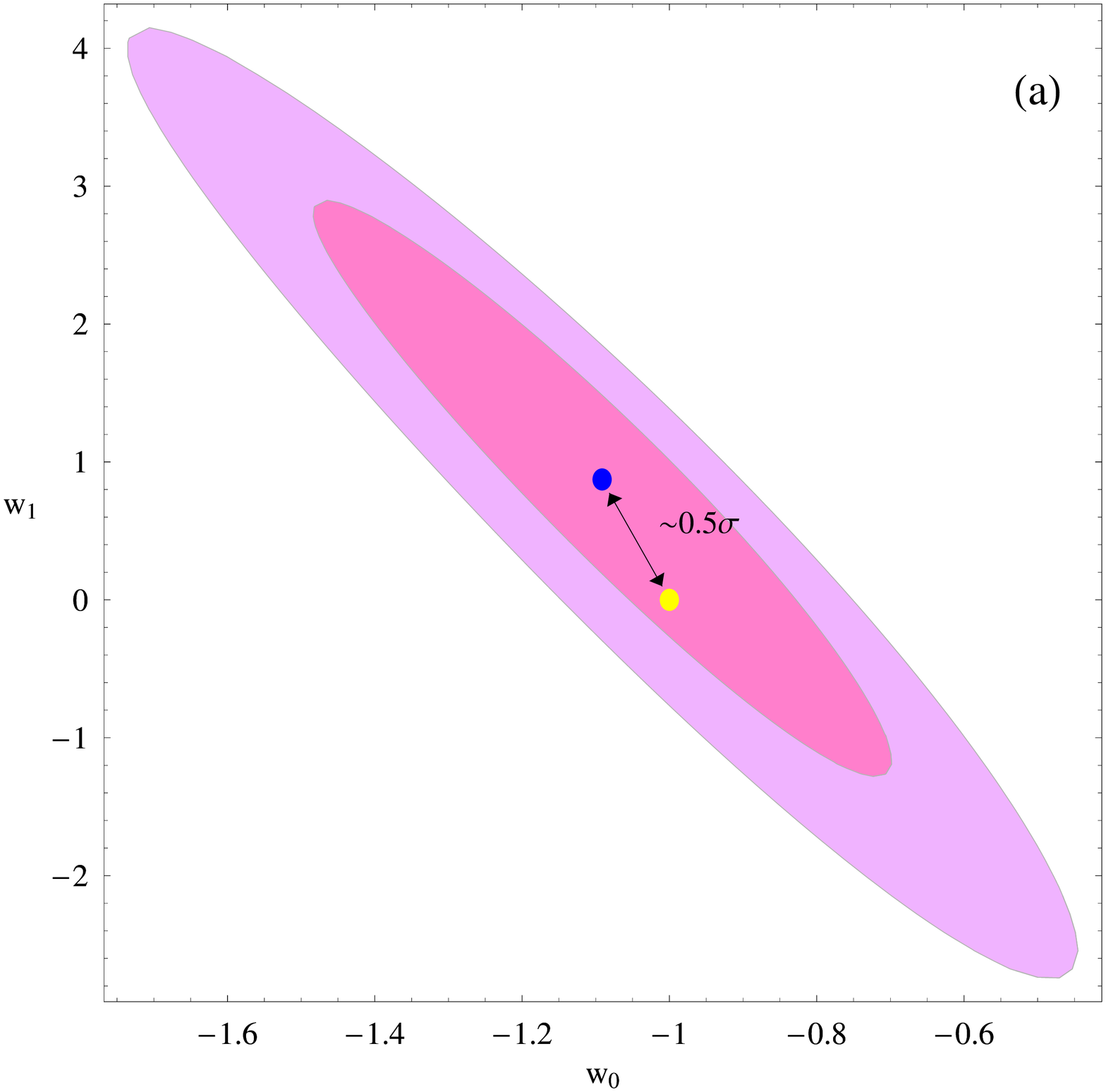}}}
\end{minipage}\hspace{2pc}%
\begin{minipage}{18pc}
\hspace{0pt}\vspace{0.2cm}\rotatebox{0}{\resizebox{1.057\textwidth}{!}{\includegraphics{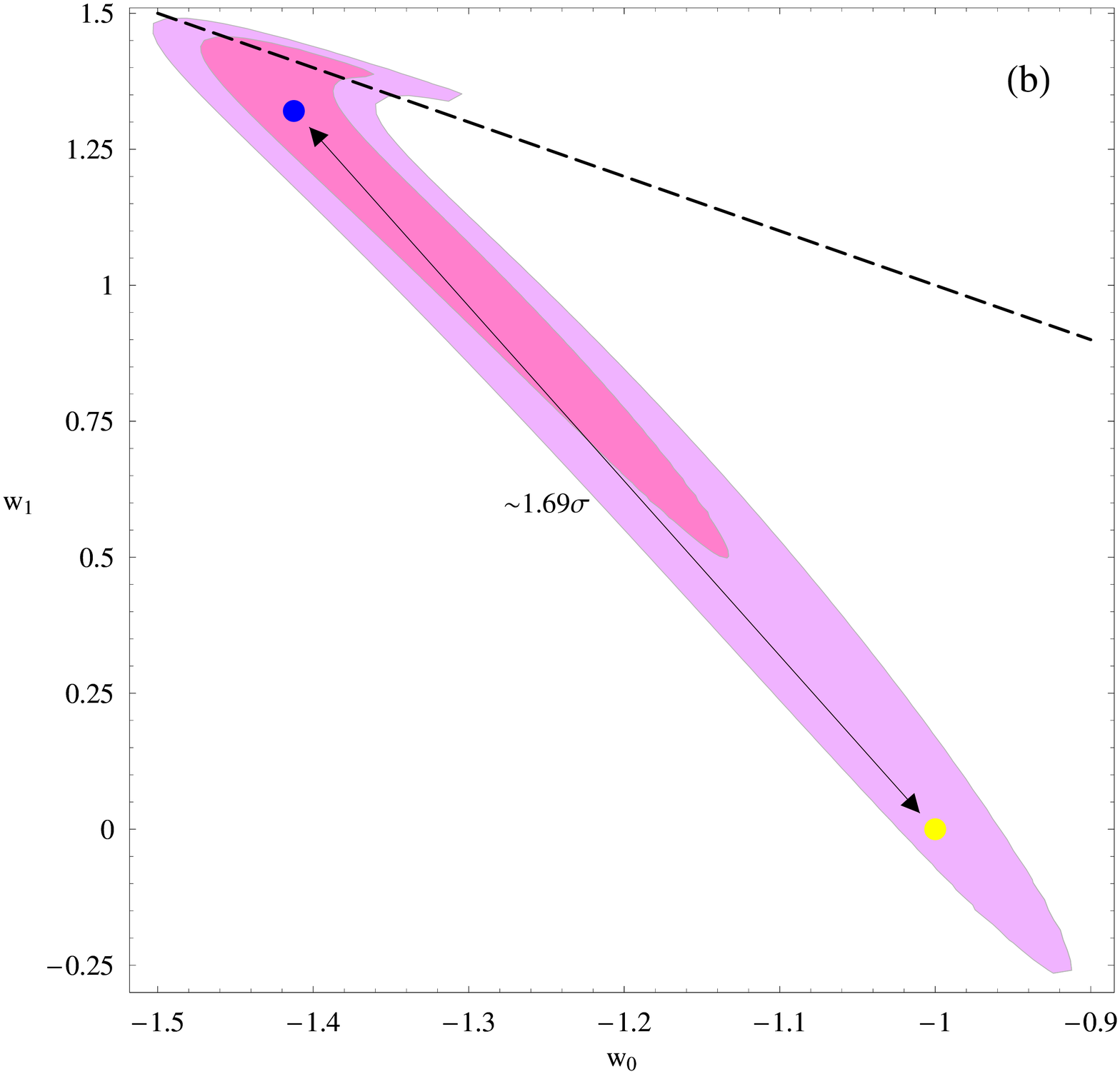}}}
\end{minipage}\hspace{2pc}%
\vspace{0pt}{\caption{The $68.3\%$ and $95.4\%$ $\chi^2$ confidence
contours in the $(w_0,w_1)$ parameter space for each dataset
category: Fig 1a for SnIa (based on data from \cite{Davis:2007na})
and Fig 1b for CMB+BAO data (based on data from
\cite{Wang:2007mza,Percival:2007yw}) for $\om=0.24$ and $\ob=0.042$
(best fits of WMAP3). The blue dots correspond to the $(w_0,w_1)$
best fit, while the yellow dots correspond to $\Lambda$CDM (-1,0).}}
\label{fig1}
\end{figure*}

\subsection{Standard Candles}
\subsubsection{SnIa}
We use the SnIa dataset of Davis et. al. \cite{Davis:2007na}
consisting of four subsets: ESSENCE \cite{WoodVasey:2007jb} (60
points), SNLS \cite{Astier:2005qq} (57 points), nearby
\cite{Riess:2004nr} (45 points) and HST \cite{Riess:2006fw} (30
points).

These observations provide the apparent magnitude $m(z)$ of the
supernovae at peak brightness after implementing the correction for
galactic extinction, the K-correction and the light curve width-luminosity
correction. The resulting apparent magnitude $m(z)$ is related to
the luminosity distance $D_L(z)$ through \be m_{th}(z)={\bar M}
(M,H_0) + 5 log_{10} (D_L (z)) \label{mdl} \ee where in a flat
cosmological model \be D_L (z)= (1+z) \int_0^z
dz'\frac{H_0}{H(z';\om,w_0,w_1)} \label{dlth1} \ee is the Hubble
free luminosity distance ($H_0 d_L$),  and ${\bar M}$ is the
magnitude zero point offset and depends on the absolute magnitude
$M$ and on the present Hubble parameter $H_0$ as \ba
{\bar M} &=& M + 5 log_{10}(\frac{H_0^{-1}}{Mpc}) + 25= \nn \\
&=& M-5log_{10}h+42.38 \label{barm}. \ea The parameter $M$ is the
absolute magnitude which is assumed to be constant after the above
mentioned corrections have been implemented in $m(z)$.

The SnIa datapoints are given, after the corrections have been
implemented, in terms of the distance modulus \be
\mu_{obs}(z_i)\equiv m_{obs}(z_i) - M \label{mug}\ee The theoretical
model parameters are determined by minimizing the quantity \be
\chi^2_{SnIa} (\om,w_0,w_1)= \sum_{i=1}^N \frac{(\mu_{obs}(z_i) -
\mu_{th}(z_i))^2}{\sigma_{\mu \; i}^2 } \label{chi2} \ee where
$N=192$ and $\sigma_{\mu \; i}^2$ are the errors due to flux
uncertainties, intrinsic dispersion of SnIa absolute magnitude and
peculiar velocity dispersion. These errors are assumed to be
Gaussian and uncorrelated. The theoretical distance modulus is
defined as \be \mu_{th}(z_i)\equiv m_{th}(z_i) - M =5 log_{10} (D_L
(z)) +\mu_0 \label{mth} \ee where \be \mu_0= 42.38 - 5 log_{10}h
\label{mu0}\ee and $\mu_{obs}$ is given by (\ref{mug}). The steps we
followed for the minimization of (\ref{chi2}) are described in
detail in Refs.
\cite{Nesseris:2004wj,Nesseris:2005ur,Nesseris:2006er}.
\begin{figure*}[!t]
\rotatebox{0}{\hspace{-3cm}\resizebox{1.3\textwidth}{!}{\includegraphics{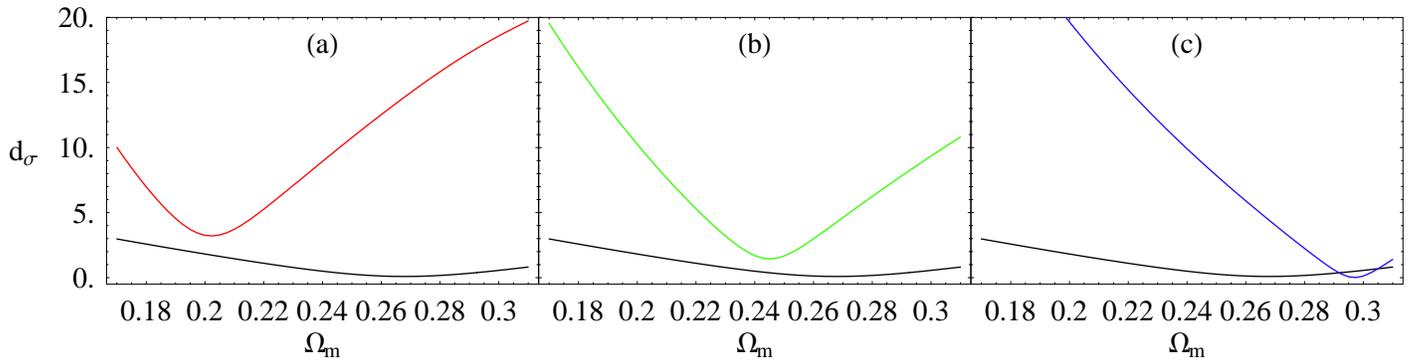}}}
\vspace{0pt}{\caption{The $\sigma$-distance of the best fit
parameters $(w_0,w_1)$ to $\Lambda$CDM (-1,0) for SnIa data (black
lines) and for CMB-BAO data (colored lines), as a function of $\om$
for $\ob=0.034$ Fig. 2a, $\ob=0.042$ Fig. 2b  and $\ob=0.049$ Fig.
2c.}} \label{fig2}
\end{figure*}

\begin{figure*}[!t]
\hspace{0cm}\rotatebox{0}{\resizebox{1\textwidth}{!}
{\includegraphics{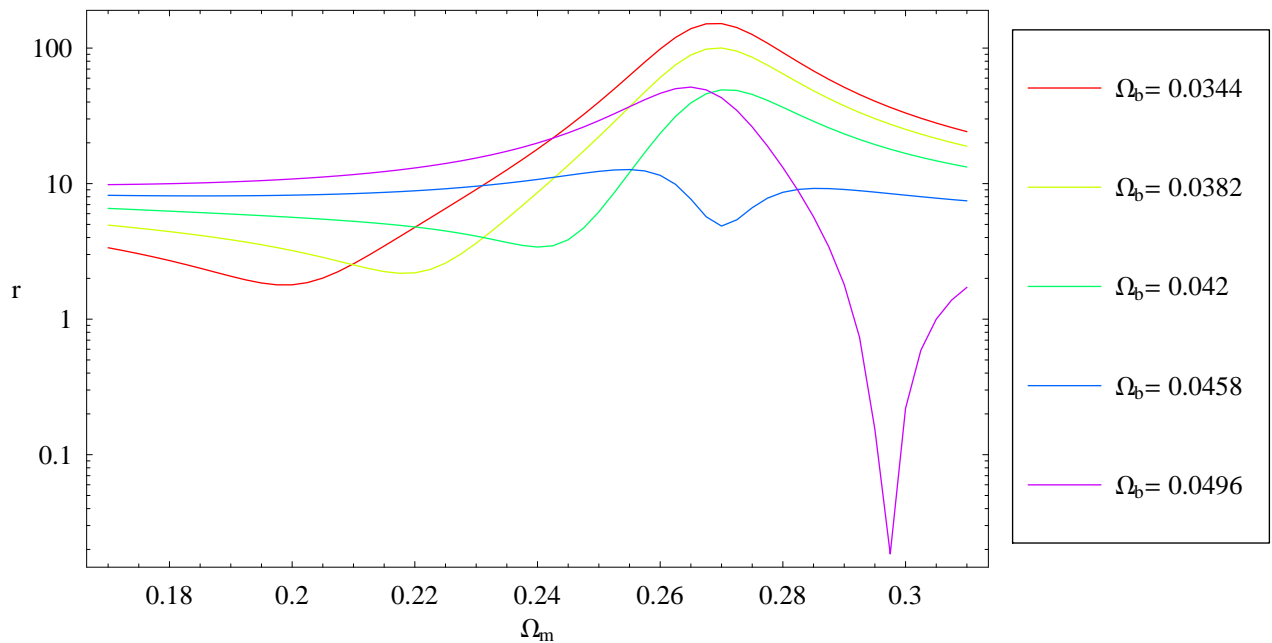}}} {\hspace{0pt}\caption{The ratios of
the $\sigma$-distances of the best fit parameters $(w_0,w_1)$ to
$\Lambda$CDM (-1,0) for SnIa standard candles over the one for
CMB-BAO standard ruler data as defined in eq.(\ref{ratio}), as a
function of $\om$ for various values of $\Omega_b$.}} \label{fig3}
\end{figure*}

\begin{figure*}[!t]
\begin{minipage}{18pc}
\hspace{0pt}\rotatebox{0}{\resizebox{1\textwidth}{!}{\includegraphics{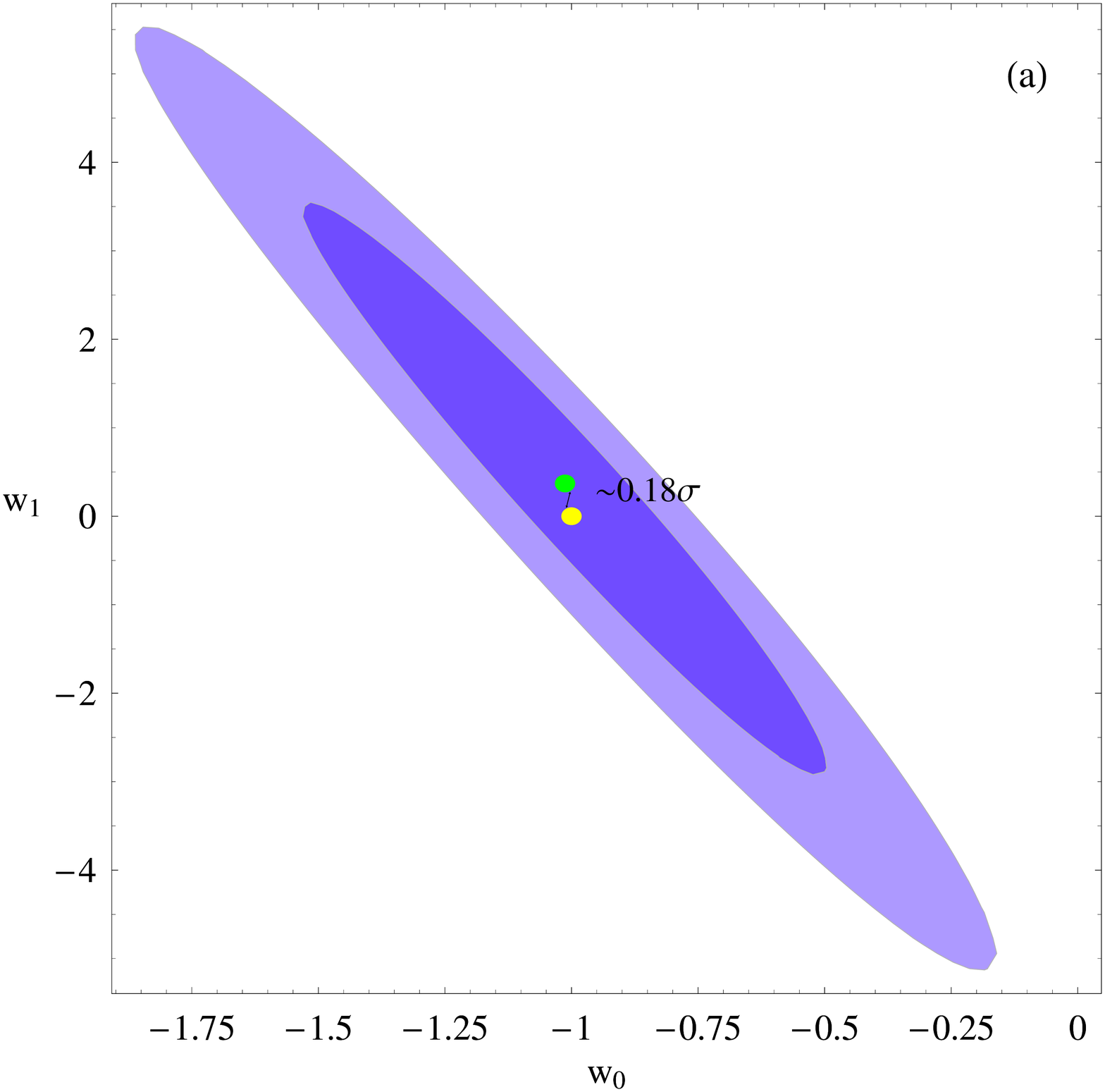}}}
\end{minipage}\hspace{4pc}%
\begin{minipage}{18pc}
\hspace{0pt}\rotatebox{0}{\resizebox{1\textwidth}{!}{\includegraphics{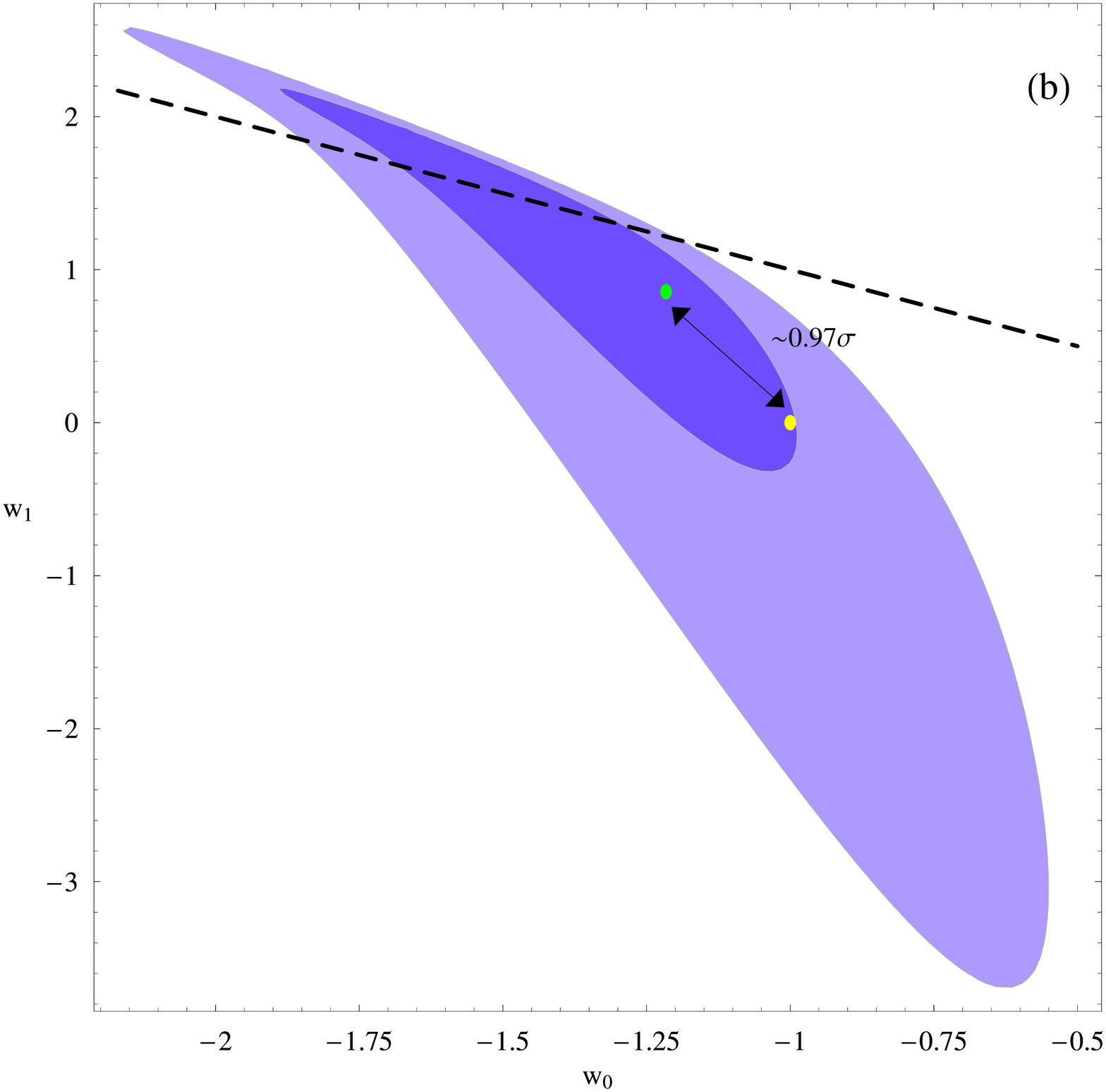}}}
\end{minipage}\hspace{4pc}%
\vspace{0pt}{\caption{The $68.3\%$ and $95.4\%$ $\chi^2$ confidence
contours in the $(w_0,w_1)$ parameter space for each of the old
dataset categories\cite{Astier:2005qq,Eisenstein:2005su} for
$\om=0.24$. The green dots correspond to the $(w_0,w_1)$ best fit,
while the yellow dots correspond to $\Lambda$CDM .}} \label{fig4}
\end{figure*}

\begin{figure*}[!t]
\begin{minipage}{18pc}
\hspace{0cm}\rotatebox{0}{\resizebox{1\textwidth}{!}{\includegraphics{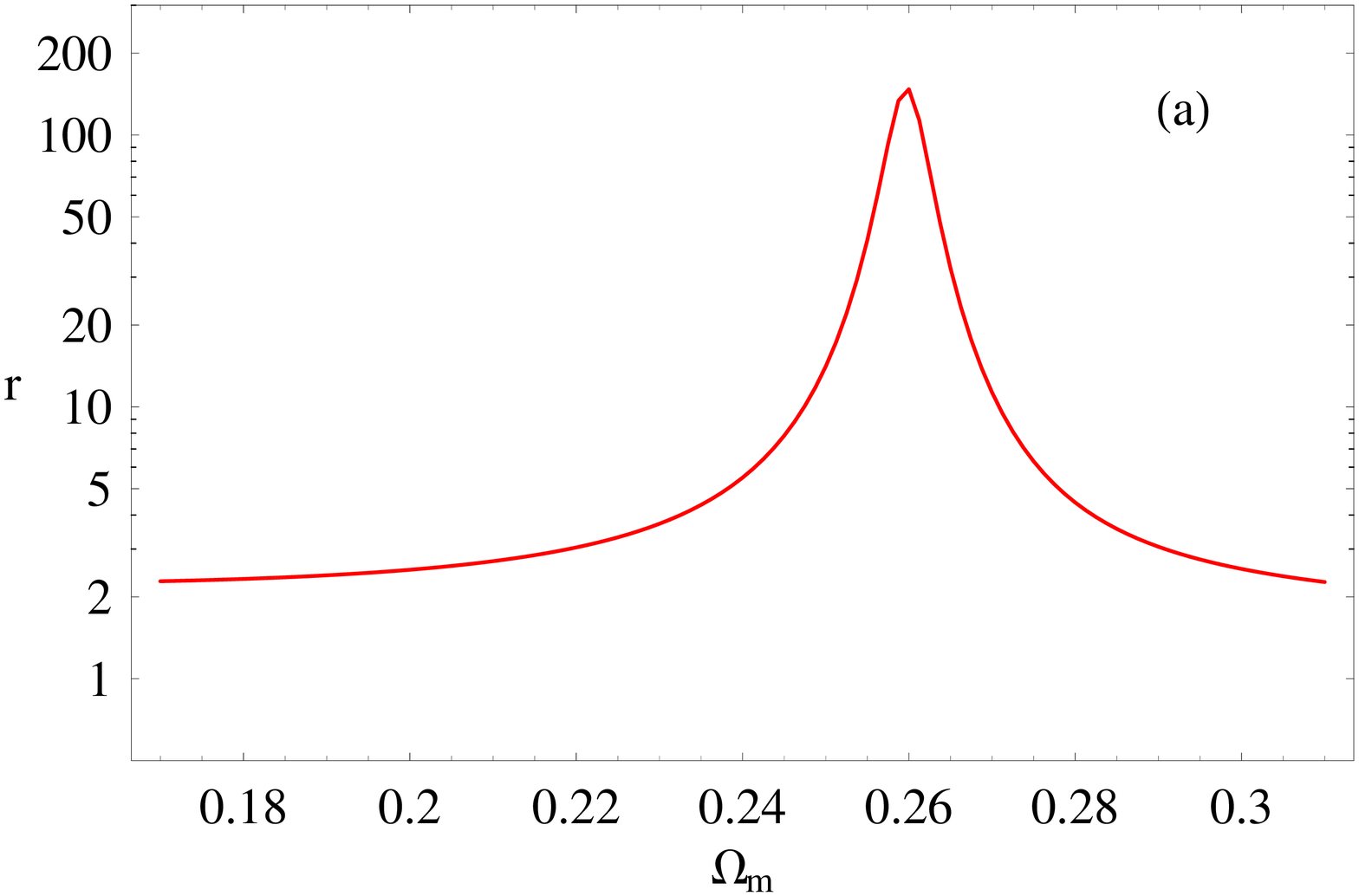}}}
\end{minipage}\hspace{4pc}
\begin{minipage}{18pc}
\hspace{0cm}\vspace{0.2cm}\rotatebox{0}{\resizebox{1\textwidth}{!}{\includegraphics{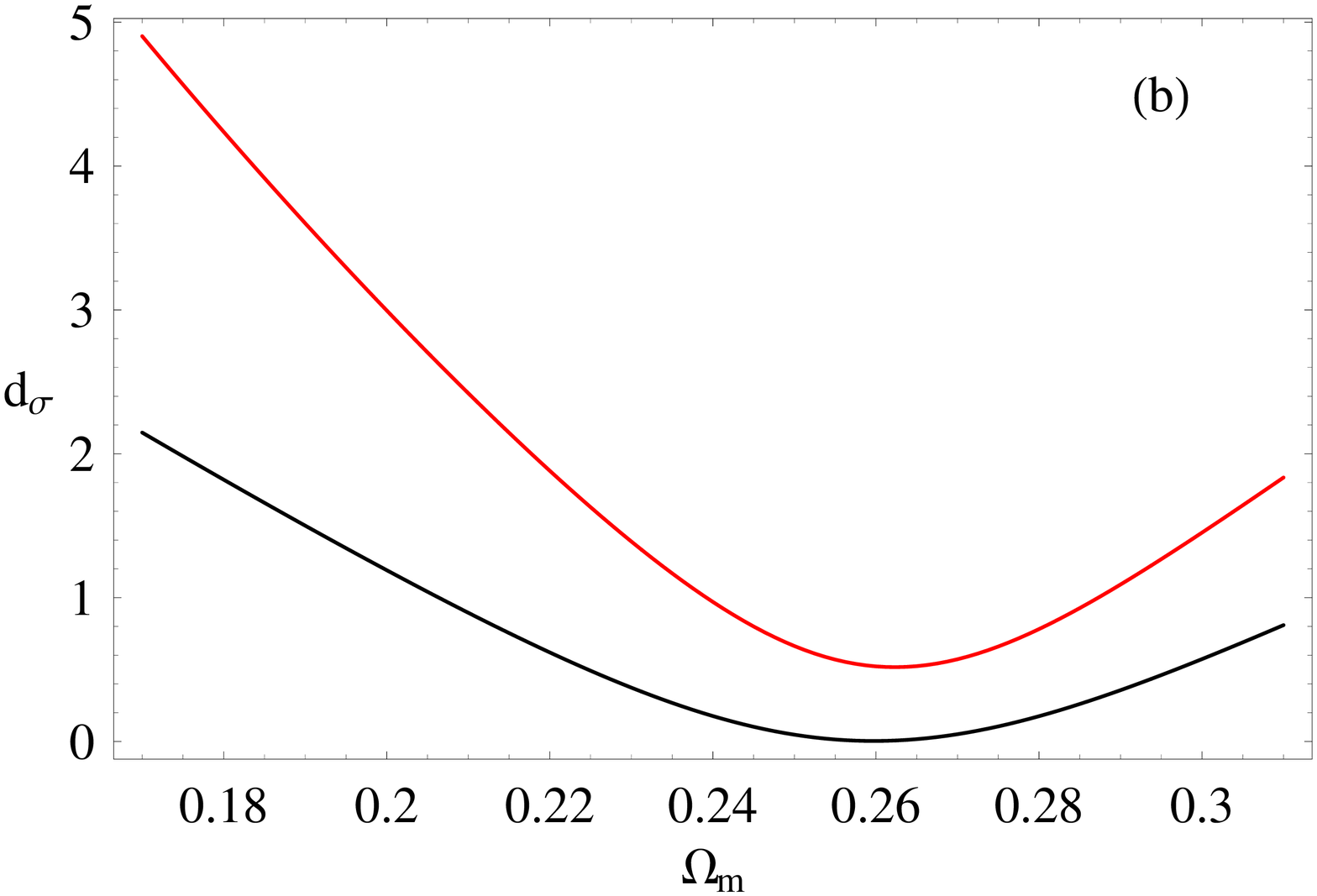}}}
\end{minipage}\hspace{4pc}
\vspace{0pt}{\caption{ (a) The ratio of the $\sigma$-distance of the
best fit parameters $(w_0,w_1)$ to $\Lambda$CDM (-1,0) for SnIa over
the one for CMB-BAO data, as a function of $\om$ for the early
CMB+BAO data\cite{Eisenstein:2005su}. (b) The $\sigma$-distance of
the best fit parameters $(w_0,w_1)$ to $\Lambda$CDM
 for SnIa data (black line) and for CMB+BAO
data (red line), as a function of $\om$.}} \label{fig5}
\end{figure*}

\subsection{Results}
We consider separately the standard ruler data ($\chi^2_{SR}\equiv
\chi^2_{CMB}+\chi^2_{BAO}$) and the standard candle data
($\chi^2_{SnIa}$), and perform minimization of the corresponding
$\chi^2$ with respect to the parameters $w_0$ and $w_1$ for various
priors of $\om$ and $\ob$ in the $2\sigma$ range of the the WMAP3
best fit ie $0.17\leq \om \leq 0.31 $, $0.034\leq \ob \leq 0.049$.

In Fig. 1 we show the $68.3\%$ and $95.4\%$ $\chi^2$ confidence
contours in the $(w_0,w_1)$ parameter space for the two dataset
categories (standard ruler and standard candle data) for $\om=0.24$
and $\ob=0.042$ (the best fit of the WMAP3 CMB data
\cite{Spergel:2006hy}). Fig. 1a shows the $(w_0,w_1)$ contours
obtained using SnIa data \cite{Davis:2007na} (standard candles)
while Fig. 1b shows the corresponding contours assuming CMB+BAO data
\cite{Wang:2007mza,Percival:2007yw} (standard rulers). The blue dots
correspond to the $(w_0,w_1)$ best fit, while the yellow dots
correspond to $\Lambda$CDM $(w_0,w_1)=(-1,0)$. The distance in units
of $\sigma$ ($\sigma$-distance $d_\sigma$) of the best fit to
$\Lambda$CDM was found by converting $\Delta \chi^2=\chi^2_{\Lambda
CDM}-\chi^2_{min}$ to $d_\sigma$ ie solving \cite{press92} \be
1-\Gamma(1,\Delta \chi^2/2)/\Gamma(1)={\rm Erf}(d_\sigma/\sqrt{2})
\label{sigmas}\ee for $d_\sigma$ ($\sigma$-distance), where $\Delta
\chi^2$ is the $\chi^2$ difference between the best-fit and
$\Lambda$CDM and ${\rm Erf}()$ is the error function. Notice that
$\Lambda$CDM is consistent at less than $1\sigma$ level according to
the SnIa data ($d_\sigma^{SnIa}\simeq 0.5$ in Fig. 1a), while the
corresponding consistency level reduces to $d_\sigma^{SR}\simeq
1.7\sigma$ for the standard ruler CMB+BAO data. This mild difference
in trends between standard candles and standard rulers persists also
for all values of $\om$ in the $2\sigma$ range of WMAP3 best fit.
This is demonstrated in Fig. 2 where we show the $\sigma$-distance
$d_{\sigma}^{SR}$ superposed with $d_{\sigma}^{SnIa}$ as a function
of $\om$  for $\ob=0.034$ (Fig. 2a), $\ob=0.042$ (Fig. 2b) and
$\ob=0.049$ (Fig. 2c). These values of $\ob$ span the $2\sigma$
range of the corresponding WMAP3 best fit. Notice that the
$\sigma$-distance between best fit values and $\Lambda$CDM values is
consistently larger when using standard ruler data (colored lines
are consistently above black lines).

An alternative way to see this trend is to plot the ratios $r$ of
the $\sigma$-distances defined as: \be r(\om)\equiv
\frac{d_{\sigma}^{SR}}{d_{\sigma}^{SnIa}} \label{ratio} \ee  These
plots are shown in Fig. 3 for five values of $\ob$ spanning the
$2\sigma$ range of WMAP3. Notice that the colored lines are
consistently above the line $r=1$ indicating that the
$\sigma$-distance is found to be consistently larger when using
standard ruler data. An exception to this rule is the case
corresponding to high values of both $\ob$ and $\om$ set to values
$2\sigma$ or more, away from their best fit (see magenta line
corresponding to $\ob=0.049$, for $\om >0.29$).

\begin{figure}[!t]
\rotatebox{0}{\hspace{0cm}\resizebox{0.5\textwidth}{!}{\includegraphics{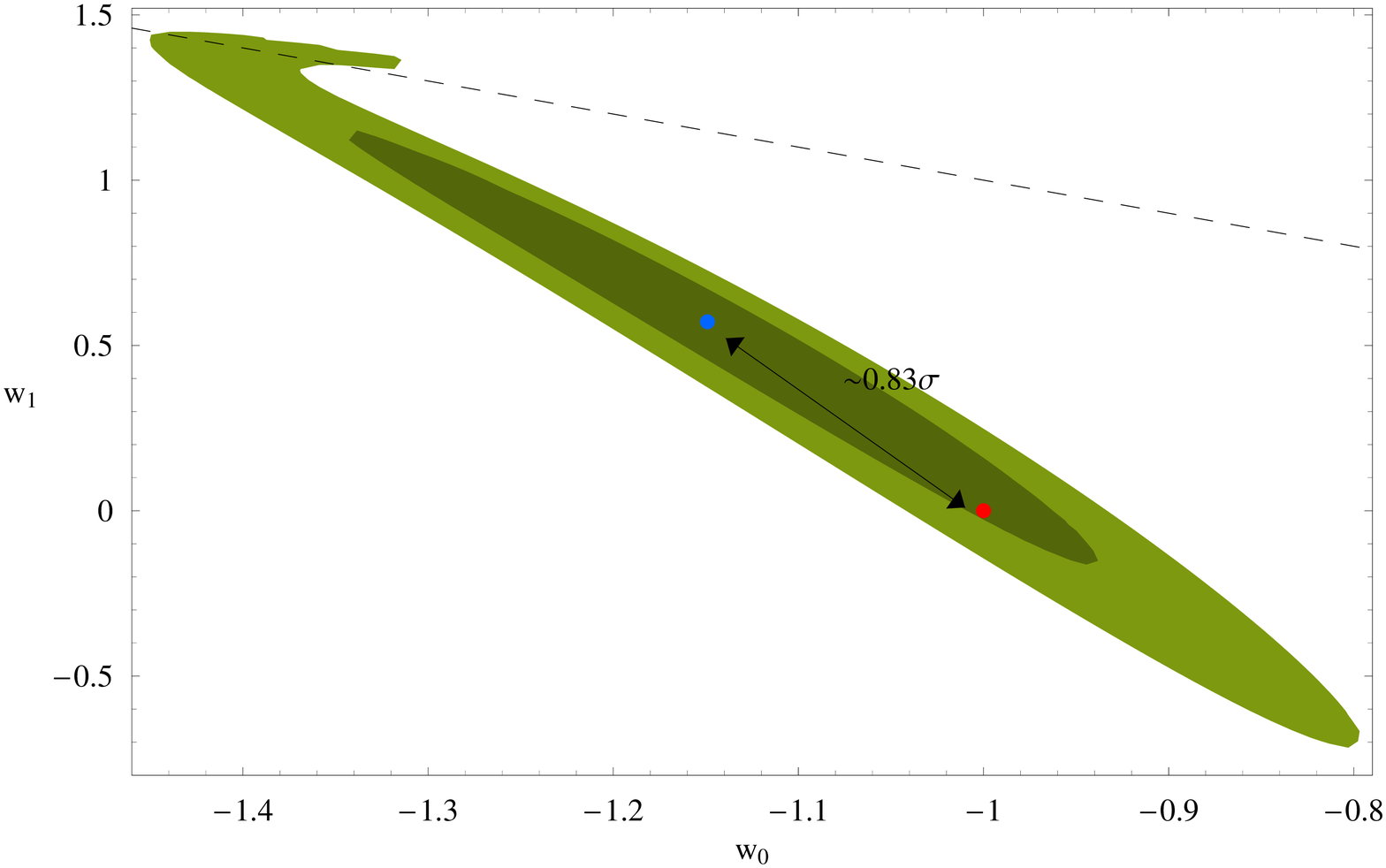}}}
{\hspace{0pt}\caption{The $68.3\%$ and $95.4\%$ $\chi^2$ confidence
contours in the $(w_0,w_1)$ parameter space for the combined
datasets SnIa+CMB+BAO for $\om=0.24$. The blue dot corresponds to
the $(w_0,w_1)$ best fit while the red dot to $\Lambda$CDM (-1,0).}}
\label{fig6}
\end{figure}
The above plots reveal a consistent trend of the standard ruler
CMB+BAO data for a mild preference for crossing of the phantom
divide line $w = -1$, while the recent SnIa data seem to favor
$\Lambda$CDM. Figs. 4 and 5 show corresponding plots obtained using
earlier data (SNLS \cite{Astier:2005qq} vs CMB+BAO
\cite{Wang:2006ts,Eisenstein:2005su} at $z = 0.35$), where a similar
consistent trend is observed.

An interesting feature of the contours of Fig. 1b is the deformation
appearing for relatively large values of $w_1$. There is a simple
way to understand this deformation. For $(w_0,w_1)$ parameter values
satisfying $w_0+w_1 \gsim 0$, the dark energy equation of state is
approximately constant and positive at early times and therefore the
corresponding dark energy density dominates over the matter density.
As a result the Hubble expansion rate is significantly modified over
the whole range from $z=0$ to $z_{CMB}$ and the corresponding
integral of the shift parameters becomes very sensitive to parameter
changes. The effect is even more significant for the shift parameter
$l_a$ which involves the sound horizon $r_s$ in the denominator (see
eq. (\ref{shiftl}). The sound horizon drops more dramatically than
the shift parameter $R$ when the dark energy dominates at early
times because the corresponding integral (\ref{rs}) depends {\it
only} on the early time behavior of the expansion rate $H(a)$. This
effect is demonstrated by plotting the $w_0+w_1=0$ line in Fig. 1b
which coincides approximately with the region where the contour
deformation starts. The same line is also plotted in Fig. 4b
corresponding to early CMB+BAO data\cite{Eisenstein:2005su} and
involving only one shift parameter ($R$). In this case the
deformation effect is milder because the z integral corresponding to
$R$ spreads over a wide range of redshifts from $z=0$ to $z=1089$
and the effect of dark energy domination is somewhat smeared out.

We also construct the likelihood contours using the combined
SnIa+CMB+BAO data for $\om=0.24$ and $\ob=0.042$ corresponding to
the best fit WMAP3 parameter values. As expected, the
$\sigma$-distance between best fit and $\Lambda$CDM is at about
$1\sigma$, ie intermediate between the standard candle and standard
ruler cases (see Fig. 6).

\section{Conclusions-Discussion}
We have demonstrated that there is a systematic difference in trends
between standard candle (SnIa) and standard ruler (CMB+BAO) data.
The former data are significantly more consistent with $\Lambda$CDM
than the later for practically all ($\om,\ob$) parameter priors
within the $2\sigma$ range of WMAP3. In fact, the standard ruler
data demonstrate a mild preference for a best fit $w(z)$ that
crosses the phantom divide line $w=-1$.

\begin{figure*}[!t]
\rotatebox{0}{\hspace{-1cm}\resizebox{1.1\textwidth}{!}{\includegraphics{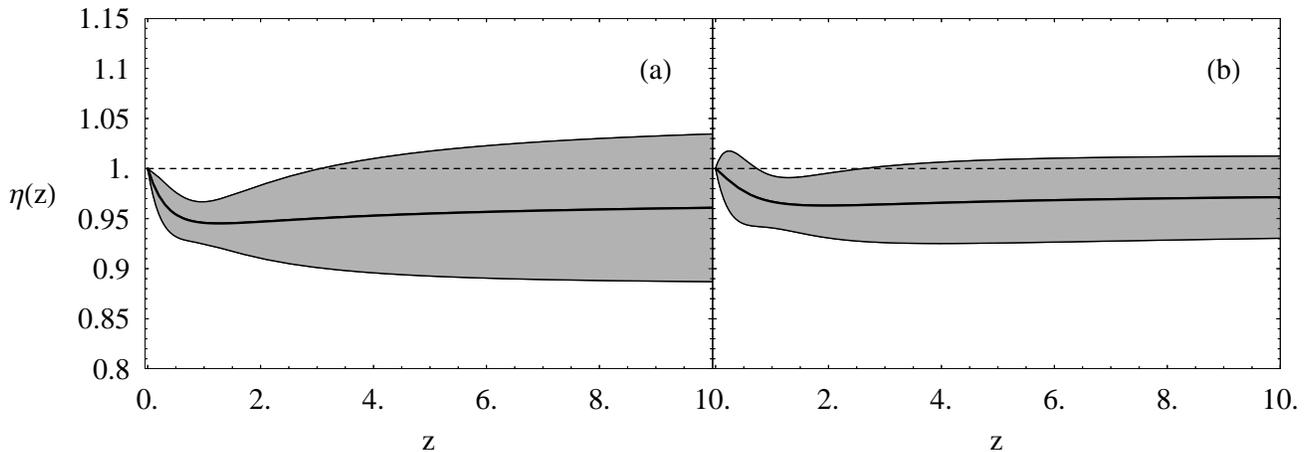}}}
{\hspace{0pt}\caption{The constrains on $\eta(z) \equiv
\frac{d_L(z)}{d_A(z) (1+z)^2}$ for $\om = 0.24$ in Fig. 7a and for
$\om=0.27$ in Fig. 7b. Clearly the anticipated value $\eta = 1$ is
within $1-2\sigma$ for both priors used.}} \label{fig7}
\end{figure*}

This systematic difference in trends can be attributed to one of the
following:
\begin{itemize}
\item {\bf Statistical Effects:} There is an ($\om,\ob$) parameter
range where both datasets are consistent with each other and with
$\Lambda$CDM at the $2\sigma$ level (see eg. Fig. 3b with $\om
\simeq 0.25$). Therefore, for these parameter values the two
datasets are consistent with each other and with $\Lambda$CDM at the
$1\sigma-2\sigma$ level and the trend we observe could well be a
statistical fluctuation.
\item {\bf Systematic-Physical Effects:} As discussed in Ref.
\cite{Bassett:2003vu} distances based on standard candles and
standard rulers should agree as long as three conditions are met:
(1) photon number is conserved, (2) gravity is described by a metric
theory and (3) photons are traveling on unique null geodesics. If at
least one of these conditions is not met then equations (\ref{dl1})
and (\ref{da1}) will lead to generically different forms for the
Hubble expansion rate $H(z)$ due to the violation of the distance
duality relation (\ref{distdual}). For example, lensing of SnIa by
compact objects, if not properly accounted for, would tend to
violate condition (3) and induce artificial brightening of distant
SnIa. Alternatively, photon number violation (due eg to photon
mixing \cite{Mirizzi:2006zy}) would lead to artificial dimming of
the SnIa.
\end{itemize}

In order to investigate the possible existence of systematic
physical effects, we have used our results to test the cosmic
distance duality relation (\ref{distdual}). In particular, we use
our results for the best fit parameter values ($(w_0,w_1)$) and
their error bars obtained from each dataset to derive constraints on
the parameter $\eta(z)$. These constraints are shown in Fig. 7a for
$\om = 0.24$ and in Fig. 7b $\om=0.27$. Clearly, the anticipated
value $\eta = 1$ is within $2\sigma$ for both priors used. Assuming
a prior of $\om=0.24$ and taking an average value for $\eta(z)$ in
the range $0<z<40$ (as for large enough $z$ $\eta(z)$ converges (see
Fig 7)), yielding the value $\bar{\eta}=0.96 \pm 0.07$ which is
within $1\sigma$ from the anticipated value $\eta = 1$. The
consistency is somewhat reduced if we average over a more recent
redshift range. In the range $1<z<2$ we find $\bar{\eta}=0.95 \pm
0.025$ which is consistent with the anticipated value $\eta=1$ at
the $2\sigma$ level. Similar results are obtained for other priors
of $\om$ within $2\sigma$ from the WMAP3 best fit. Therefore,
despite the mild difference in trends between SnIa standard candles
and CMB+BAO standard rulers, we find no statistically significant
evidence for violation of the distance duality relation.

An interesting extension of this work would be the inclusion of more
 data from both categories. For example gamma ray
bursts \cite{Meszaros:2006rc} could also be included as standard
candles and X-ray profiles of clusters \cite{Uzan:2004my} or radio
galaxies \cite{Daly:2007pp} could be included as standard rulers in
order to investigate if the mild difference in trends we have
identified, persists in more general categories of data.

The Mathematica files with the numerical analysis of the paper can
be found at http://leandros.physics.uoi.gr/rulcand/rulcand.htm

\section*{Acknowledgements}
We thank Y. Wang, P. Mukherjee, W. Percival and E. Majerotto for useful
discussions. This work was supported by the European Research and
Training Network MRTPN-CT-2006 035863-1 (UniverseNet), by the
University of the Basque Country through research grant
GIU06/37, and by the Spanish Ministry of Education and
Culture through research grant FIS2004-01626. S.N.
acknowledges support from the Greek State Scholarships Foundation
(I.K.Y.).

\end{document}